\documentclass[aps,prl,tightenlines,twocolumn,showpacs,preprintnumbers,amsmath,amssymb]{revtex4}

\usepackage{graphicx}
\usepackage{dcolumn}
\usepackage{bm}
\usepackage{epsfig}

\begin{document}

\title{ Comment: Essence of intrinsic tunneling: Distinguishing intrinsic features from artifacts [Phys.Rev.B 72 (2005) 094503]. }

\author{V.M.Krasnov}

\affiliation { Department of Physics, Stockholm University,
AlbaNova University Center, SE-10691 Stockholm, Sweden}


\begin{abstract}

In a recent paper V.Zavaritsky has argued that interlayer ($c-$
axis) current-voltage characteristics of high temperature
superconductors (HTSC) are Ohmic, and has claimed to disprove all
findings of intrinsic tunnelling spectroscopy, as well as
existence of the interlayer tunnelling and the intrinsic Josephson
effect in HTSC, as such.

In this comment I demonstrate, that the genuine interlayer
current-voltage characteristics are strongly non-Ohmic, which
undermines the basic postulate and the logical construction of the
criticized paper.

\end{abstract}


\maketitle

Mobile charge carriers in strongly anisotropic high temperature
superconductors (HTSC), such as Bi$_2$Sr$_2$CaCu$_2$O$_{8+\delta}$
(Bi-2212), are confined in CuO$_2$ planes. The out-of-plane
$c-$axis transport in those compounds is caused by interlayer
tunnelling \cite{theory}, which results in non-metallic behavior
\cite{Timusk,Ando,Watanabe} and leads to appearance at $T<Tc$ of
the "intrinsic" Josephson effect \cite{Kleiner} between
neighboring CuO$_2$ planes. At present all major fingerprints of
the intrinsic Josephson effect were observed, including Fiske
\cite{Fiske,LatyshPhC,Kim} and Shapiro \cite{LatyshPhC,Wang} steps
in Current-Voltage characteristics (IVC's); the Josephson plasma
resonance \cite{Plasma}; thermal activation \cite{Fluct} and
macroscopic quantum tunnelling \cite{MQT} from the Josephson
washboard potential; and the flux quantization
\cite{Fiske,LatyshPhC,Kim,Latysh,Ooi}. The latter experiments
explicitly confirmed the correspondence between the stacking
periodicity of intrinsic Josephson junctions (IJJ's) and the
crystallographic unit cell of Bi-2212. The interlayer tunnelling
was also successfully employed for intrinsic tunnelling
spectroscopy, which is unique in it's ability to probe bulk phonon
\cite{Schlen} and quasiparticle
\cite{Latysh,Kras_T,Suzuki,Kras_H,Doping,Shibauchi,Lee} spectra of
HTSC.

However, in a recent paper \cite{ZavarPRB} and a series of other
publications \cite{Zavar} V.Zavaritsky denies the existence of
interlayer tunnelling in HTSC and speculates that all the
non-linear features in $c-$axis IVC's are artifacts of
self-heating.

The two basic postulates of Refs.\cite{ZavarPRB,Zavar} are:

1. That the intrinsic Josephson effect does not exist,

2. That the genuine, self-heating free $c-$axis IVC's of HTSC are
Ohmic.

The latter statement together with the assumption of
semiconducting $T-$dependence of the Ohmic resistance leads in the
presence of self-heating to non-linear IVC's, which according to
Ref. \cite{ZavarPRB} provides both qualitative and quantitative
description of all key findings of interlayer transport
experiments in HTSC, proves that all the non-linear features of
interlayer IVC's are artifacts of self-heating and disproves all
findings of intrinsic tunnelling spectroscopy.

The incorrectness of the first postulate in the face of overwhelming
and unambiguous experimental evidence
\cite{Kleiner,Fiske,LatyshPhC,Kim,Wang,Plasma,Fluct,MQT,Latysh,Ooi,Schlen}
does not require extra comments, except that it demonstrates an
obvious failure of the reviewing process in several scientific
journals.

In this comment I demonstrate incorrectness of the second
postulate. It is shown that the genuine interlayer IVC's of
Bi-2212 are strongly non-Ohmic and are represented by perfectly
periodic, multibranch IVC's, which are not affected by
self-heating. The huge, two order of magnitude, genuine
non-linearity of IVC's represents the extent of irrelevance of the
model advocated by Zavaritsky to the essence of intrinsic
tunnelling in HTSC and undermines the logical construction of Ref.
\cite{ZavarPRB}.

Intrinsic tunnelling characteristics of HTSC mesa structures
exhibit a characteristic multi-branch structure due to one-by one
switching of IJJ's from the superconducting into the resistive
state \cite{Kleiner}. Each time a new junction is switched into
the resistive state the dissipation power within the mesa
increases and the effective temperature of the mesa rises as a
result of self-heating \cite{Heating}:

\begin{equation}
\Delta T = PR_{th},  \label{Eq.Zav}
\end{equation}

where $P=VI$ is the total dissipation power and $R_{th}$ is the
effective thermal resistance of the mesa, which depend on the
sample geometry and temperature \cite{Heating,Insitu}. The
progressive self-heating with the branch number may result in a
systematic distortion of the IVC in the way, shown in Fig. 1.
Namely, the critical current and the separation between branches
decrease with the branch number. At large $P$, significant
self-heating is indicated by progressive back-bending of the
branches.

Fig. 1 shows how the intrinsic IVC is distorted by extreme
self-heating. It shows an IVC at the base temperature $T_0=4.2K$
for a large Pb-doped Bi-2212 mesa structure, containing $N\simeq
40$ IJJ's. The IVC is measured in the four-probe configuration.
The extremely large $P>5 mW$ at the end of the multi-branch
structure in the IVC is caused first of all by a very large
critical current density $J_c \sim 10^4 A/cm^2$ in this Pb-doped
Bi-2212 crystal. A Similar distortion of intrinsic IVC's at
comparable $P$ can be seen in Fig. 10 a) of Ref.\cite{Schlen},
Fig. 1 of Ref.\cite{Johnson}

\begin{figure}
\noindent
\begin{minipage}{0.48\textwidth}
\epsfxsize=0.9\hsize \centerline{ \epsfbox{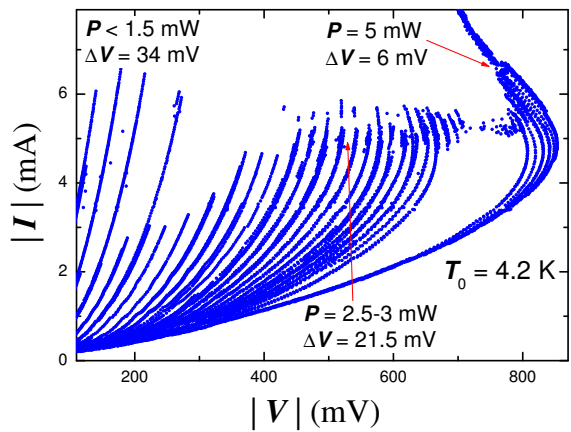}}
\caption{( color ) Superimposed positive and negative parts of the
IVC of a large Pb-doped Bi-2212 mesa at $T_0=4.2K$. It
demonstrates how the IVC is distorted by self-heating: the
critical current and the separation between branches decrease with
the branch number due to progressive self-heating. At very large
dissipation power significant self-heating is indicated by
progressive back-bending of the branches.}\label{Fig.1}
\end{minipage}
\end{figure}

Fig. 2 shows an opposite example of a small self-heating. In Fig.
2 a) the IVC of a small underdoped Bi-2212 mesa with a small
$J_c<400 A/cm^2$ \cite{Doping} and approximately the same number
of IJJ's, $N=34$, is shown. Properties of this mesa were studied
in Refs. \cite{Fluct,Insitu}. A combination of small area and
$J_c$ results in a two order of magnitude smaller $P$ at the end
of the multi-branch structure, point A in Fig. 2 a), than at the
similar point in Fig. 1. Thin lines in Fig. 2 a) represent the
multiple integer of the last branch divided by $N=34$, which
indicates perfect periodicity of the branches, typical for
interlayer IVC's of Bi-2212 mesas \cite{Schlen,Kras_T,Wang,Fluct}.
Fig. 2 b) shows voltage jumps, $\Delta V$, between the branches as
a function of the branch number. It is seen that $\Delta V$ is
independent of the branch number. Small deviations from the mean
$\Delta V$ are caused by thermal fluctuations of the switching
current \cite{Fluct,MQT}.

\begin{figure}
\noindent
\begin{minipage}{0.48\textwidth}
\epsfxsize=0.9\hsize \centerline{ \epsfbox{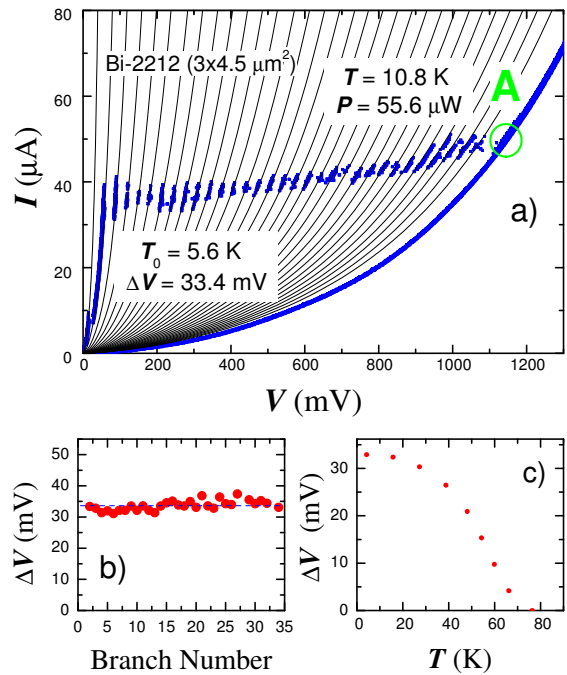}}
\caption{(color) a) The IVC of a small underdoped Bi-2212 mesa at
$T_0=5.6K$. Thin lines are multiple integers of the last branch
divided by $N=34$ and demonstrate perfect periodicity of the
branches. Panels b) and c) show the voltage jumps between branches
as a function of the branch number and the base temperature,
respectively.}\label{Fig.2}
\end{minipage}
\end{figure}

The separation between branches provides an independent test for
the extent of self-heating, because $\Delta V$ depends on $T$, as
shown in Fig. 2 c). From Fig. 2 c) it follows that branches can
remain periodic only if the successive increase of $T$ with branch
number does not exceed $\sim 15K$. This is consistent with in-situ
measured $T=10.8K$  at $P=55.6 \mu W$ point A in the IVC from Fig.
2 a) \cite{Insitu}. Similarly, for the case of extreme
self-heating shown in Fig. 1, we may conclude that the mesa is
heated to $T \simeq 60 K$ at the end of the multi-branch structure
at $P \simeq 5 mW$, again consistent with the measured
self-heating $\Delta T/P \sim 10 K/mW$ at $T = 60 K$
\cite{Insitu}. Furthermore, atomic separation between IJJ's in the
mesa leads to uniform self-heating within the mesa, see
Appendix-A.

Appendix-B presents a detailed comparison of intrinsic IVC's with
the self-heating model of Ref. \cite{ZavarPRB}, which, in contrast
to claims of Ref. \cite{ZavarPRB}, demonstrates a severe
contradiction between experimental data and the model. On the
other hand, this failure is obvious already from the raw data in
Figs. 1,2. This data speaks for itself: The distorted periodicity
of branches in Fig. 1 indicates significant self-heating. To the
contrary, the undistorted perfect periodicity of branches in Fig.
2 a) implies that the shape of the branches is not affected by
self-heating. However, $V/I$ changes by approximately two orders
of magnitude in this voltage range.

Thus, observation of the periodic strongly non-Ohmic branches in
interlayer IVC's can only mean that:

i) the self-heating along the branches is negligible;

ii) those periodic branches represent the genuine self-heating
free IVC's of Bi-2212;

iii) The genuine interlayer IVC's are strongly non-linear.

The latter statement is the main conclusion of this comment. It
disproves the basic postulate of Ref.\cite{ZavarPRB} and
undermines the logical construction of that paper. From Fig. 2 and
a large collection of similar data reported in literature, it can
be concluded that the genuine non-linearity of intrinsic IVC's
exceeds two orders of magnitude at low $T$, which also represents
the degree of irrelevance of the model suggested in the criticized
paper to the essence of intrinsic tunnelling in HTSC.

\section{APPENDIX A}

The author of Ref. \cite{ZavarPRB} speculates about "peculiar
temperature distribution along the sample". However,
$T-$distribution within the mesa can be easily estimated, without
making any assumptions about heat flow mechanism outside the mesa,
or dissipation mechanism inside the mesa.

Fig. 3 shows the schematics of heat flow in Bi-2212 mesa
structure. The heat $P=VI$, produced in the mesa, can flow down to
the crystal and upwards into the contact electrode, characterized
by heat resistances $R_{cr}$ and $R_{el}$, respectively. $R_{1,2}$
represent heat resistances of top and bottom parts of the mesa.
The total self-heating is

\begin{equation}
\Delta T = P R_{th}= P
\frac{(R_1+R_{el})(R_2+R_{cr})}{R_1+R_2+R_{el}+R_{cr}},
\label{Eq.3}
\end{equation}
where $R_{th}$ is the effective thermal resistance of the sample,
which according to Ref. \cite{ZavarPRB} is $R_{th} = (Ah)^{-1}
\simeq 4 \times 10^4 K/W$. The one-dimensional nature of heat flow
within the mesa allows a straightforward estimation:

\begin{equation}
R_{1,2}=\frac{sN_{1,2}}{A\kappa_c},\label{Eq.2}
\end{equation}
where $s=15.5 \AA$ is the interlayer spacing, $N_{1,2}$ is the
number of layers in top/bottom parts of the mesa, and $\kappa_c$
is the $c-$axis thermal conductivity. For Bi-2212, $\kappa_c \sim
0.5 W/K m$ at $T= 30K$ \cite{Uher}. For the mesa with $A= 10
\times 10 \mu m^2$, the thermal resistance per layer $R_1(N=1)
\simeq 31 K/W$ is three orders of magnitude smaller than $R_{th}$.
In two extreme cases, $R_{cr} \gg R_{el}$ and $R_{cr} = R_{el}$,
the ratios of temperature difference per junction in the mesa to
the total self-heating of the mesa are $R_1/R_{th} \simeq 7.8
\times 10^{-4}$ and $R_1/2R_{th} \simeq 3.9 \times 10^{-4}$,
respectively.

For the mesa with $N=10$ and $P = 1 mW$ (note that $P < 0.06 mW$
at point A in the IVC of Fig. 2 a)) the maximum temperature
difference within the mesa, $\Delta T_{mesa}$ is only $\sim 0.3
K$.

Furthermore, this is a strongly overestimated value because: (i)
$\kappa_c$ measured in Ref.\cite{Uher} was limited by stacking
faults in large Bi-2212 crystals, which are absent in our mesas.
The pure phononic thermal conductivity is expected to be almost
isotropic, which would imply that the actual $\kappa_c$ in the
mesa is close to $\kappa_{ab}$, i.e., eight times higher. (ii) The
estimation was obtained for the case of heat diffusion within the
mesa. In reality, the c-axis heat transport in Bi-2212 is
predominantly phononic and ballistic at the atomic scale
\cite{Insitu,Uher}. This will considerably reduce $\Delta
T_{mesa}$ for mesas with the total hight less then the phononic
mean free path, i.e. for $N \lesssim 1000$.

Therefore, for typical mesas all layers within the mesa are heated
uniformly, irrespective of where exactly within the mesa the
dissipation takes place.

\begin{figure} 
  \includegraphics[width=0.4\textwidth]{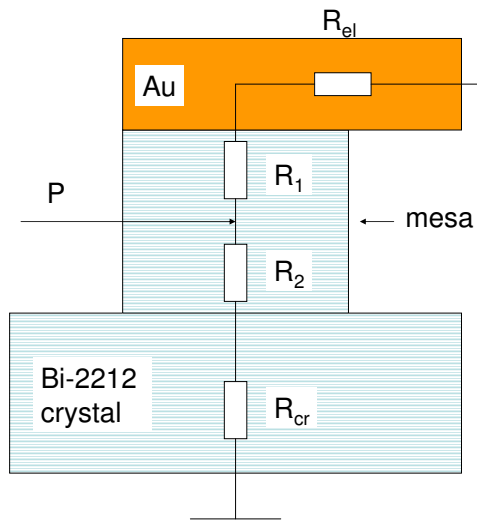}\\
  \caption{(color) Schematics of heat flow from the mesa
structure.}\label{Fig.3}
\end{figure}

\section{APPENDIX B}

One of the critical assumptions of Ref.\cite{ZavarPRB} is that the
semiconducting zero-bias resistance $R_0(T)$ represents $V/I$ at
finite bias along the IVC's. This a direct consequence of the
second postulate, see above. A seeming matching between decaying
$V/I(P)$ and $R_0(T)$ is demonstrated in Fig.1 of Ref.
\cite{ZavarPRB} for some IVC, which, however, does not bear any
resemblance with classical multibranch interlayer characteristics
\cite{Schlen,Kras_T,Wang,Fluct}. Below I will repeat the same
fitting for the case of conventional intrinsic IVC's.

Fig. 4 represents $V/I(P)$ (left and top axes) for the IVC from
Fig.2 a). The IVC in full scale is shown in inset of Fig. 4. The
decaying part of the curve represents the resistance of the last
branch, while the linearly increasing part represents multiple
branches in the IVC. The dashed and solid lines in Fig. 4
represent $R_0 (T)$ measured with a small ac-modulation current
and the quasiparticle resistance with all $N=34$ IJJ's in the
resistive state $R_0^{N} (T)$ obtained from numerical
differentiation of the IVC's, respectively (left and bottom axes).
On the first glance, a good match between $V/I(P)$ and $R_0(T)$,
in a limited $T-$range, can be obtained if we allow an arbitrary
offset and stretching along the $T-$axis of $R_0(T)$. However,
several inconsistencies can be seen in such a "fit":

\begin{figure}
\noindent
\begin{minipage}{0.48\textwidth}
\epsfxsize=0.9\hsize \centerline{ \epsfbox{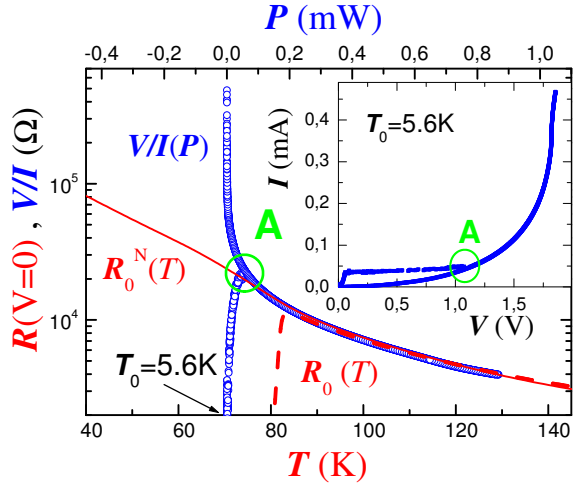}}
\caption{(color) The main panel shows the IVC from the inset, for
the same mesa as in Fig.2, replotted as $V/I$ vs. $P=VI$ (top
axis). Dashed and solid lines represent the zero bias resistance
$R_0$ and the quasiparticle resistance $R_0^{34}$ vs. $T$ (bottom
axis), respectively. $R_0(T)$ is matched to $V/I(P)$ using the
procedure suggested in Ref.\cite{ZavarPRB}. Inset shows the
original $I-V$ characteristics of a Bi-2212 mesa at $T_0=5.6K$.
Major deviations between $V/I(P)$ and $R_0(T)$ at low $P$ and a
huge offset $T(P=0) \simeq 70 K \gg T_0=5.6 K$, indicated by an
arrow, reveal severe contradiction between experimental data and
the self-heating model of Ref. \cite{ZavarPRB}. }\label{Fig.4}
\end{minipage}
\end{figure}

i) $T-$offset, required for matching, leads to $T(P=0) \simeq 70 K
\gg T_0=5.6 K$, as indicated by the arrow in Fig. 4. A similar
offset can be seen in Fig. 1 of Ref. \cite{ZavarPRB}. Therefore,
such a "fit" simply does not make any sense. The seeming
coincidence in Fig. 4 indicates simply that two smooth decaying
functions can be matched in the limited interval using two fitting
parameters.

ii) major deviations of $V/I(P)$ and $R_0(T)$ occur at low $P$,
where the main, almost 2-order of magnitude, drop of $V/I$ occurs.
The severe discrepancy between $V/I(P)$ and $R_0(T)$ at low bias
indicates that the IVC is non-linear in this bias range.

Strictly speaking, the scaling between $V/I(P)$ and $R_0(T)$
should not be expected even within the self-heating model because
the relation between $T$ and $P$ in Eq.(\ref{Eq.Zav}) is
non-linear due to $T-$dependence of $R_{th}$ \cite{Insitu}. One
should instead expect self-scaling of $V/I(P)$ curves at different
$T_0$, using the offset $\Delta P$ as the only fitting parameter,
representing the power required to compensate the difference in
$T_0$ between different measurements. Such scaling does not
require constant $R_{th}$ and just implies that equal $P$ would
heat the mesa by the same $\Delta T$, irrespective of how the mesa
reached the initial temperature.

\begin{figure}
\noindent
\begin{minipage}{0.48\textwidth}
\epsfxsize=0.9\hsize \leftline{ \epsfbox{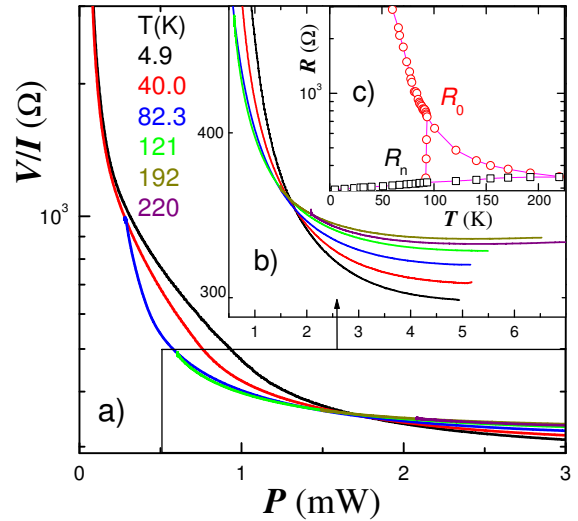}}
\caption{(color) Check for a self-scaling of $V/I(P)$ curves: a) a
set of $V/I (P=VI)$ characteristics for another Bi-2212 mesa at
different $T_0$. The offset $\Delta P$ was used to fit the origin
of each curve at $P=0$ to the nearest curve with lower $T_0$. b)
Enlarged high $P$ part of a). c) Temperature dependencies of $R_0$
(circles) and high bias resistance $R_n$ (squares) for the same
mesa. It is seen that $R_n$ is smaller than $R_0$ at any $T$.}
\label{Fig.5}
\end{minipage}
\end{figure}

Fig. 5 shows the check for such a self-scaling for the optimally
doped mesa ($T_c=93 K$, $A=3.5\times 7.5 \mu m^2$, $N=10$),
studied in Ref. \cite{Kras_T}. Fig. 5 a) shows a set of $V/I(P)$
curves at six $T_0$ from 4.9 to 220 K. For each curve the offset
$\Delta P$ was chosen in such a way that the initial point
$V/I(P=0)$ would coincide with $V/I(\Delta P)$ of the nearest
curve with lower $T_0$. Apparently, the self-scaling is non
existing for conventional intrinsic IVC's. Fig. 5 b) shows zoom-in
of the high power part of the plot. Here self-heating should be
the largest. It is seen that the IVC's asymptotically approach
some "normal" resistance $R_n$ \cite{Kras_T}. Experimental values
of $R_n$ and $R_0$ vs. $T_0$ are shown in Fig. 5 c). Comparison of
$R_n$ and $R_0$ shows that $R_n$ is smaller than $R_0$ at any $T$.
This makes the fitting of $V/I$ by $R_0(T)$ technically
impossible, indicating that the IVC's have to be non-Ohmic.


\begin{references}

\bibitem{theory} S.Chakravarty and P.W.Anderson, Phys. Rev. Lett. 72 (1994) 3859; D.G.Clarke, S.P.Strong, and P.W.Anderson,
Phys. Rev. Lett. 74 (1995) 4499; W.Kim and J.P.Carbotte, Phys.
Rev. B 63 (2000) 054526; N.Shah and A.J.Millis, Phys. Rev. B 65
(2001) 024506.

\bibitem{Timusk} A.V.Puchkov, D.N.Basov and T.Timusk, J.Phys.Cond
Matt. 8 (1996) 10049

\bibitem{Ando}Y.Ando, G.S.Boebinger, A.Passner, N.L.Wang, C.Geibel and F.Steglich, Phys. Rev. Lett. {\bf 77} (1996) 2065

\bibitem{Watanabe} T. Watanabe, T. Fujii, and A. Matsuda, Phys. Rev. Lett. {\bf 84} (2000) 5848

\bibitem{Kleiner} R.Kleiner and P.M\"{u}ller, Phys.Rev.B {\bf 49}, 1327 (1994)

\bibitem{Fiske} V.M.Krasnov, N.Mros, A.Yurgens, and D.Winkler, Phys. Rev. B {\bf 59}, 8463 (1999)

\bibitem{LatyshPhC} Yu.I.Latyshev, A.E.Koshelev, V.N.Pavlenko, M.B.Gaifulin, T.Yamashita and
Y.Matsuda, Physica C 367 (2002) 365

\bibitem{Kim} S.M.Kim, H.B.Wang, T.Hatano, S.Urayama, S.Kawakami, M.Nagao, Y.Takano, T.Yamashita, and K.Lee, Phys. Rev.
B 72 (2005) 140504(R)

\bibitem{Wang} H.B. Wang, P.H.Wu, and T.Yamashita, Phys. Rev. Lett. {\bf 87}, 107002 (2001)

\bibitem{Plasma} K.Lee, W.Wang, I.Iguchi, M.Tachiki, K.Hirata, and
T.Mochiku, Phys. Rev. B 61 (2000) 3616; M.B.Gaifullin, Y.Matsuda,
N.Chikumoto, J.Shimoyama, K.Kishio, and R.Yoshizaki, Physica C 362
(2001) 228.

\bibitem{Fluct} V.M.Krasnov, T.Bauch and P.Delsing, Phys. Rev.
B 72 (2005) 012512

\bibitem{MQT} K.Inomata, S.Sato, K.Nakajima, A.Tanaka,
Y.Takano, H.B.Wang, M.Nagao, H.Hatano, and S.Kawabata, Phys. Rev.
Lett. 95 (2005) 107005

\bibitem{Latysh} Yu.I.Latyshev, S.J.Kim, V.N.Pavlenko, T.Yamashita, and L.N.Bulaevskii, Physica C {\bf 362} (2001) 156

\bibitem{Ooi} S.Ooi, T.Mochiku, and K.Hirata, Phys. Rev. Lett. 89
(2002) 247002

\bibitem{Schlen} K.Schlenga, R.Kleiner, G.Hechtfischer,
M.M\"{o}{\ss}le, S.Schmitt, P.M\"{u}ller, Ch.Helm, Ch.Preis,
F.Forsthofer, J.Keller, H.L.Johnson, M.Veith, and E.Steinbei\ss,
Phys. Rev. B 57 (1998) 14518

\bibitem{Kras_T} V.M. Krasnov, A.Yurgens, D.Winkler, P.Delsing, and T.Claeson, Phys. Rev. Lett. {\bf 84},
5860 (2000)

\bibitem{Suzuki} M.Suzuki and T.Watanabe, Phys. Rev. Lett. {\bf 85}, 4787 (2000)

\bibitem{Kras_H} V.M. Krasnov, A.E.Kovalev, A.Yurgens and D.Winkler, Phys. Rev. Lett. {\bf 86}, 2657 (2001)

\bibitem{Doping} V.M.Krasnov, Phys. Rev. B {\bf 65}, 140504(R) (2002)

\bibitem{Shibauchi} L. Krusin-Elbaum, T. Shibauchi, and C.H. Mielke, Phys. Rev. Lett. {\bf
92} (2004) 097005

\bibitem{Lee} M.H. Bae, J.H. Choi, H.J. Lee, and K.S. Park,
cond-mat/0512664

\bibitem{ZavarPRB} V.N. Zavaritsky, Phys. Rev. B {\bf 72} (2005) 094503

\bibitem{Zavar} V.N. Zavaritsky, Physica C {\bf 404} (2004), 440; Phys. Rev. Lett. {\bf 92} (2004) 259701

\bibitem{Heating} V.M.Krasnov, A.Yurgens, D.Winkler and P.Delsing, J. Appl. Phys. {\bf 89}, 5578 (2001); {\em ibid.} {\bf
93}, 1329 (2003); V.M.Krasnov, Physica C {\bf 372-376}, 103
(2002);

\bibitem{Insitu} V.M.Krasnov, M.Sandberg and I.Zogaj, Phys. Rev. Lett. {\bf 94} (2005)
077003;

\bibitem{Johnson} H.L.Johnson, G.Hechtfischer, G.G\"{o}tz,
R.Kleiner and P.M\"{u}ller, J.Appl.Phys. 82 (1997) 756

\bibitem{Uher} M.F.Crommie and A.Zettl, Phys.Rev.B {\bf43} (1991)
408. 


\end{references}
\end{document}